\documentstyle[12pt]{article}
\def\Journal#1#2#3#4{{#1} {\bf #2}, #3 (#4)}

\def\NPB{{\em Nucl. Phys.} B}
\def\PLB{{\em Phys. Lett.}  B}

\def\PRD{{\em Phys. Rev.} D}

\def\be{\begin{equation}}
\def\ee{\end{equation}}
\def\bea{\begin{eqnarray}}
\def\eea{\end{eqnarray}}

\begin{document}

\rightline{FTUV/99-28}

\vspace{1.5cm}

\begin{center}
{\large{\bf CP, T VIOLATION IN NEUTRINO OSCILLATIONS}}
\end{center}

\vspace{1.5cm}

\begin{center}
J. BERNAB\'EU
\end{center}

\vspace{1cm}

\begin{center}
{\it Department of Theoretical Physics, University of
Valencia} 
\end{center}

\vspace{2cm}

\begin{abstract}
{\small{
The manifestation of CP, T violation
in the leptonic sector is studied for flavour neutrino oscillations,
both in vacuum and in matter. Different conditions of short-base-line
versus long-base-line experiments are discussed.}}
\end{abstract}

\newpage

\section{Introduction}

Complex neutrino mixing for 3 family Dirac neutrinos leads to
CP and T violation effects in neutrino oscillations \cite{UNO}. In
view of the vigorous experimental programme in this field, the
study of CP violation becomes an interesting topic.

The neutrino states of definite flavour $\alpha$, as generated
by well defined weak interaction properties, are related to 
neutrino states of definite mass $m_k$ by

\begin{equation}
\nu_\alpha = \sum_k U_{\alpha k} \nu_k
\label{eq:uno}
\end{equation}

\noindent
where $U$ is the unitary mixing matrix which, for 3 families, 
depends on 3 mixing angles and  1 CP phase.

If the "$\alpha$" state is prepared at $t = 0$, the probability 
amplitude that, at time $t$, it is manifested as the "$\beta$" state is

\begin{equation}
A (\alpha \rightarrow \beta ; t) = \sum_k U_{\alpha k} U_{\beta k}^*
exp[- i E_k t]
\label{eq:dos}
\end{equation}

We observe that the time-dependent amplitude contains the interference
of different "$k$" terms, with different weak phases in $U_{\alpha k}
\, U_{\beta k}^*$ and different oscillation phases governed by $E_k$.
These ingredients are necessary and sufficient to generate CP violation
in the oscillation probability.

In Section 2 we discuss the CPT-invariance condition, together with
the CP-odd and T-odd asymmetries. In Section 3 the CP-asymmetry for
3 family neutrino oscillation is considered and the conditions for a 
non-vanishing value are obtained. These results will lead to the need
of long-base-line (LBL) experiments for CP studies. In Section 4 the 
CP-odd asymmetry is built in this case for hierarchical neutrino masses.
These LBL experiments have to include, however, matter effects and
in Section 5 we show that these matter effects are large and they 
constitute an undesired background for CP violation effects. Due to
this fake phenomenon, Section 6 studies T-odd asymmetries which are
free from this problem. Section 7 answers the question related  to
the possible Majorana character of neutrinos. Section 8 summarizes
some conclusions and the outlook.

\section{CPT, CP, T}

>From Eq.~(\ref{eq:dos}) the requirement of CPT invariance leads
to the amplitude for conjugate flavour states

\begin{equation}
A (\bar{\alpha} \rightarrow \bar{\beta}; t) = \sum_k U_{\alpha k}^*
U_{\beta k} \, exp[-i E_k t]
\label{eq:tres}
\end{equation}

\noindent
so that we obtain the condition

\begin{equation}
CPT \Rightarrow A (\bar{\alpha} \rightarrow \bar{\beta}; t)
= A^* (\alpha \rightarrow \beta; - t) 
\label{eq:cuatro} 
\end{equation}

Eq.~(\ref{eq:cuatro}) will be assummed through this work.

CP-invariance is the statement that the probabilities for the 
original transition and for its conjugate are equal, i.e.,

\begin{equation}
CP \Rightarrow |A (\alpha \rightarrow \beta; t)|^2 =
|A (\bar{\alpha} \rightarrow \bar{\beta}; t)|^2
\label{eq:cinco}
\end{equation} 

T-invariance is the statement that the probabilities for the
original transition and for its inverse are equal, i.e,

\begin{equation}
\begin{array}{ll}
T \Rightarrow & | A (\alpha \rightarrow \beta ; t)|^2 = | A (\beta
\rightarrow \alpha; t)|^2\\
& |A (\bar{\alpha} \rightarrow \bar{\beta}; t)|^2 =
|A (\bar{\beta} \rightarrow \bar{\alpha}; t)|^2
\end{array}
\label{eq:seis}
\end{equation} 

From
these results, we have the corollaries:

i) CP, T Violation effects, i.e, the violation of Eq.~(\ref{eq:cinco}) ,
Eq.~(\ref{eq:seis}), respectively, can take place in Appearance
Experiments only. For Disappearance experiments, $\beta = \alpha$,
Eq.~(\ref{eq:seis}) is automatic and Eq.~(\ref{eq:dos}) implies

\begin{equation}
A^* (\alpha \rightarrow \alpha ; t) = A (\alpha \rightarrow \alpha;
- t)
\label{eq:siete}
\end{equation}

The combination of Eq.~(\ref{eq:siete}) and  Eq.~(\ref{eq:cuatro})
leads to the verification of
Eq.~(\ref{eq:cinco}), q.e.d.
As a consequence, no CP or T violation effect can be manifested in reactor
or solar neutrino experiments.

ii) The (numerator of) CP-odd Asymmetry is given by

\begin{equation}
D_{\alpha \beta} \equiv |A (\alpha \rightarrow \beta; t)|^2
- |A (\bar{\alpha} \rightarrow \bar{\beta}; t)|^2
\label{eq:ocho}
\end{equation}

CPT-invariance implies $D_{\alpha \beta} = - D_{\beta \alpha}$
and the use of Eq.~(\ref{eq:dos}) and the Unitarity of the
Mixing Matrix implies $\sum_{\beta \neq \alpha} D_{\alpha \beta} 
= 0$. These constraints lead to a unique $D$ for 3 flavours:

\begin{equation}
D_{e \mu} = D_{\mu \tau} = D_{\tau e}
\label{eq:nueve}
\end{equation}

iii) The (numerators of) T-odd Asymmetries are given by

\begin{equation}
\begin{array}{l}
T_{\alpha \beta} \equiv | A (\alpha \rightarrow \beta; t)|^2
- | A (\beta \rightarrow \alpha ; t)|^2 = | A
(\alpha \rightarrow \beta ; t)|^2 - | A (\alpha \rightarrow \beta;
-t)|^2 \\
\bar{T}_{\alpha \beta} \equiv | A (\bar{\alpha} \rightarrow \bar{\beta}
; t)|^2
- | A (\bar{\beta} \rightarrow \bar{\alpha} ; t)|^2 = | A
(\bar{\alpha} \rightarrow \bar{\beta} ; t)|^2 - | 
A (\bar{\alpha} \rightarrow \bar{\beta};
-t)|^2 \\[2ex]
\end{array}
\label{eq:diez}
\end{equation}

\noindent
where use of Eq.~(\ref{eq:dos}) has been made in the right-hand
side.  Eq.~(\ref{eq:diez}) leads to the important conclusion that
\underline{$T_{\alpha \beta}, \bar{T}_{\alpha \beta}$ are odd functions
of time}. One should be aware that   Eq.~(\ref{eq:diez}) needs
an hermitian Hamiltonian for the evolution of the system. In fact,
the above conclusion is not valid for the $K^0 \bar{K}^0$ system.

\section{CP-Asymmetry}

The numerator Eq.~(\ref{eq:ocho}) of the CP-odd asymmetries can be 
calculated using Eq.~(\ref{eq:dos}) for the amplitudes. In the limit
of ultrarelativistic neutrinos, one obtains

\begin{equation}
D_{\alpha \beta} = \sum_{k >j} I_{\alpha \beta; j k} \sin
\frac{\Delta m_{kj}^2 L}{ 2 E}
\label{eq:once}
\end{equation}

\noindent
where $L \simeq t$ is the distance between the source and the detector,
$E$ is the neutrino energy and $\Delta m_{kj}^2 \equiv m_k^2
- m_j^2$. The I's containing mixing angles and the CP phase are given
by

\begin{equation}
I_{\alpha \beta ; jk} = 4 Im [U_{\alpha j} U_{\beta j}^*
U_{\alpha k}^* U_{\beta k} ]
\label{eq:doce}
\end{equation}

\noindent
which show the rephasing invariance of the observables explicitly.

Suppose that only the highest $m^2$-value is relevant for the
neutrino oscillation experiment, assuming a hierarchy in neutrino 
masses. This statement, which can be considered as the definition of a 
short-base-line (SBL) experiment, means that the approximations

\begin{equation}
\left.
\begin{array}{c}
\Delta m^2 \simeq \Delta m_{31}^2 \simeq \Delta m^2_{32}\\
\frac{\Delta m^2_{21}}{2 E} L << 1
\end{array}
\right\}
\label{eq:trece}
\end{equation}

\noindent
are fulfilled. In the limit of neglecting terms of order $\frac{
\Delta  m_{21}^2}{2 E} L$, the asymmetry Eq.~(\ref{eq:once}) 
becomes

\begin{equation}
D_{\alpha \beta}^{(SBL)} \simeq (I_{\alpha \beta ; 13}
+ I_{\alpha \beta ; 23}) \sin \frac{\Delta m^2L}{2E}
\label{eq:catorce}
\end{equation}

\noindent
which vanishes due to the cyclic character of the I's: 
$I_{\alpha \beta; 23} = 
I_{\alpha \beta; 31} = 
- I_{\alpha \beta; 13}$.

The lesson learnt from this limit is immediate: the 3 families have to
participate ACTIVELY in order to generate a non-vanishing CP-odd 
observable. It is not enough to know that there are 3 
non-degenerate neutrinos in Nature
and the presence of mixing among all of them.  $\Delta m_{21}^2$ has
to participate.
Furthermore, in order to generate a non-vanishing value for the I's
one needs ALL the mixing angles and the unique CP phase different from
zero \cite{DOS}.

One thus concludes that a significant CP-odd asymmetry needs the
consideration of neutrino oscillations in long-base-line (LBL) 
experiments \cite{TRES,CUATRO,CINCO,SEIS}. The meaning of this
requirement is that both $\Delta m^2$'s, i.e., $\Delta m_{31}^2 \simeq
\Delta m_{32}^2$ and $\Delta m_{21}^2$, have to be accessible.

Two comments should be considered to soften the above conclusion: i)
one can choose to keep terms of order $\frac{\Delta m_{21}^2}{2E} L
<<1$ at the expense to search for appearance transitions with very
low probability and enhance the CP-odd ratio which defines the
asymmetry; ii) even if the value of $D_{\alpha \beta}$, and thus
of the CP-odd asymmetry, vanishes under the conditions leading to
Eq.~(\ref{eq:catorce}), the existence of a non-vanishing CP-phase
can be inferred from CP-conserving observables if enough
probabilities are measured. To have information on both $P (\nu_\mu
\rightarrow \nu_\tau)$ and $P (\nu_e \rightarrow \nu_\tau)$ under
controllable conditions, one probably needs the 
neutrino facility based on 
muon-storage-rings \cite{SIETE}.

\section{CP effects in LBL experiments}

Contrary to the conditions discussed before in 
Eq.~(\ref{eq:trece}), we assume in this Section that $L/E$ is
such that

\begin{equation}
\left. \begin{array}{c}
\frac{\Delta m_{31}^2 L}{2E} \sim \frac{\Delta m_{32}^2 L}{2 E}
>> 1\\
\frac{\Delta m_{21}^2 L}{2 E} \sim 1 
\end{array}  \right\} \label{quince}
\end{equation}

The calculation of the appearance probabilities $\alpha \rightarrow \beta$
then gives

\begin{equation}
\begin{array}{l}
P_{\nu_\alpha \rightarrow \nu_\beta} = |
U^*_{\beta 1} U_{\alpha 1} + U^*_{\beta 2} U_{\alpha 2} 
exp (-i \frac{\Delta m_{21}^2 L}{2E}) |^2 + |U_{\beta 3}|^2
|U_{\alpha 3}|^2\\
 P_{\bar{\nu}_\alpha \rightarrow \bar{\nu}_\beta} = |
U_{\beta 1} U^*_{\alpha 1} + U_{\beta 2} U^*_{\alpha 2} 
exp (-i \frac{\Delta m_{21}^2 L}{2E}) |^2 + |U_{\beta 3}|^2
|U_{\alpha 3}|^2
\end{array} \label{eq:dieciseis}
\end{equation}

One notices that the heaviest neutrino contributes to these 
probabilities only through mixing without any oscillation: this
term is CP-even. What is relevant for the CP asymmetry is the
interference of two amplitudes $k = 1,2$ with diferent  weak phases
 and different oscillation phases: in going to the CP transformed
transition, the weak phase changes its sign whereas the oscillation
phase remains the same.

The difference of the two probabilities Eq.~(\ref{eq:dieciseis})
gives a CP-odd asymmetry

\begin{equation}
\left. \begin{array}{l}
D_{\alpha \beta}^{(LBL)} \simeq I_{\alpha \beta} \sin
\frac{\Delta m_{21}^2 L}{2 E}\\
I_{\alpha \beta} \equiv I_{\alpha \beta; 12} = 4 Im [U_{\alpha 1}
U_{\beta 1}^* U_{\alpha 2}^* U_{\beta 2}] 
\end{array} \right\}
\label{eq:diecisiete}
\end{equation}

It is immediate to realize that, for the 3-neutrino case, one has

\begin{equation}
I_{e \mu} = I_{\mu \tau} = I_{\tau e}
\end{equation}

 Bilenky et al. \cite{SEIS} have used present exclusion plots
for $\mu \rightarrow e$ and $\mu \rightarrow \tau $ transitions, together
with amplitude and unitarity bounds, to find allowed values for
$|I_{e \mu}|$ and $|I_{\mu \tau}|$. Maximum values of $10^{-2}$ for
$|I_{e \mu|}$ and around $10^{-1}$ for $|I_{\mu \tau}|$ are accessible.

\section{Matter effects in LBL experiments}

LBL experiments, with source and detector at the earth surface, imply
that neutrinos cross the earth in their  travel. It is mandatory to
discuss the matter effect \cite{OCHO,NUEVE,DIEZ}

The effective Hamiltonians for neutrinos and antineutrinos are given
in the flavour basis for 3 families:

\begin{equation}
\begin{array}{l}
H_\nu = \frac{1}{2E} \left\{ U \left( \begin{array}{lll}
m_1^2 && \\
& m^2_2 &\\
&& m_3^3 \end{array} \right) U^+ + \left(
\begin{array}{lll}
a &&\\
& 0 & \\
& & 0 \end{array} \right) \right\}\\[2ex]
H_{\bar{\nu}} = \frac{1}{2E} \left\{ U^* \left( \begin{array}{lll}
m_1^2 && \\
& m^2_2 &\\
&& m_3^2 \end{array} \right) U^T - \left(
\begin{array}{lll}
a &&\\
& 0 & \\
& & 0 \end{array} \right) \right\}
\end{array}
\label{eq:diecinueve}
\end{equation}

\noindent
where the matter effect for constant density is given by the forward
charged current interaction amplitude with electrons

\begin{equation}
a = G \sqrt{2} N_e 2 E \simeq 2.3 \times 10^{-4} eV^2
(\frac{\rho}{3 gcm^{-3}}) (\frac{E}{GeV})
\label{eq:veinte}
\end{equation}

\noindent
and $N_e$ is the (number) density of electrons. It is interesting
to build the dimensionless quantity

\begin{equation}
\frac{a L}{2 E} \simeq 0.58 \times 10^{-3} (\frac{L}{km})
\label{eq:veintiuno}
\end{equation}

\noindent
which is E-independent! This is in contrast to the E-dependent oscillation
quantity which, for $\Delta m^2$ in the range of the atmospheric neutrino
oscillation solution, gives

\begin{equation}
\frac{\Delta m^2 L}{2E} \simeq 0.75 \times 10^{-2} (\frac{L}{km} )
(\frac{GeV}{E}) 
\label{eq:veintidos}
\end{equation}

For neutrino beams with energy $E \sim 10 \, GeV$, as envisaged
by the FermiLab and CERN LBL experiments, the two quantities 
Eq.~(\ref{eq:veintiuno}) and  Eq.~(\ref{eq:veintidos}) are
comparable and one concludes that large matter effects are expected.

The diagonalization of $H_\nu$ and $H_{\bar{\nu}}$ by unitary 
matrices $U'$ and $\bar{U}'$ leads to different
eigenvalues $\tilde{m}_\nu^2$ and $\tilde{m}_{\bar{\nu}}^2$, respectively,

\begin{equation}
H_\nu = U' \frac{\tilde{m}_\nu^2}{2E} U'^+ \quad , \quad
H_{\bar{\nu}} = \bar{U}^{'*} \frac{\tilde{m}_{\bar{\nu}}^2}{2E}
\bar{U}'^T
\label{eq:veintitres}
\end{equation}

One notes that the matter effect $a \neq 0$  provokes fake CP and
CPT violation associated with the interaction with the asymmetric
medium: even the simplest diagonal probability equality for 
$\nu_e$ and $\bar{\nu}_e$ is violated, $P_{\nu_e \rightarrow \nu_e}
\neq P_{\bar{\nu}_e \rightarrow \bar{\nu}_e}$!

With the neutrino mixing in the medium described by $U'$
and $\bar{U}'$, the CP violating I's of Eq. ~(\ref{eq:doce})
are replaced by

\begin{equation}
I_{\alpha \beta ; j k}  \Rightarrow \left\{
\begin{array}{l}
I'_{\alpha \beta ; jk} = 4 Im [U'_{\alpha j } U^{'*}_{\beta j}
 U^{'*}_{\alpha k} U'_{\beta k} ]\\
\tilde{I}'_{\alpha \beta ; jk} = 4 Im [\bar{U}'_{\alpha j}
\bar{U}^{'*}_{\beta j} \bar{U}^{'*}_{\alpha k}
\bar{U}'_{\beta k} ]
\end{array} \right.
\label{eq:veinticuatro}
\end{equation}

\noindent
It is possible to prove 
\cite{SEIS}  that a real diagonal matter term, as
dictated by the Standard Model in Eq.~(\ref{eq:diecinueve}), implies
the necessary and sufficient condition

\begin{equation}
I_{\alpha \beta; j k} = 0 \Longleftrightarrow I'_{\alpha \beta ;
jk} = \bar{I}'_{\alpha \beta; jk} = 0
\label{eq:veinticinco}
\end{equation}

This  means that the identification of non-vanishing $I',  \bar{I}'$
in  matter is still a true signal of CP-violation in Nature. The CP-odd
asymmetries contain, however, additional terms which are an  undesired
background

\begin{equation}
\begin{array}{c}
P_{\nu_\alpha \rightarrow \nu_\beta} = \sum_k |U'_{\beta k}|^2 |U'_{\alpha
k}|^2 + 2 \sum_{k>j} Re [U'_{\alpha j} U^{'*}_{\beta j}
U^{'*}_{\alpha k} U'_{\beta k}] \cos \frac{\Delta \tilde{m}^2_{\nu k j}}{2E}
L\\
+  \frac{1}{2} \sum_{k > j} I'_{\alpha \beta ;jk} \sin 
\frac{\Delta \tilde{m}^2_{\nu k j}}{2E} L\\
P_{\bar{\nu}_\alpha \rightarrow \bar{\nu}_\beta} = \sum_k 
|\bar{U}'_{\beta k}|^2 |\bar{U}'_{\alpha
k}|^2 + 2 \sum_{k>j} Re [\bar{U}'_{\alpha j} \bar{U}^{'*}_{\beta j}
\bar{U}^{'*}_{\alpha k} 
\bar{U}'_{\beta k}] \cos \frac{\Delta \tilde{m}^2_{\bar{\nu} k j}}{2E}
L\\
- \frac{1}{2} \sum_{k > j} \bar{I}'_{\alpha \beta;jk} \sin 
\frac{\Delta \tilde{m}^2_{\bar{\nu} k j}}{2E} L
\end{array}
\label{eq:veintiseis}
\end{equation}

>From Eq.~(\ref{eq:veintiseis}) it is clear that, in matter,
the transition probabilitites of neutrinos and antineutrinos are
different even if CP is conserved, i.e., for $I' = \bar{I}' = 0$
one has $D_{\alpha \beta} \neq 0$. One would need the explicit
separation of the odd functions $\sin \frac{\Delta \tilde{m}^2_{kj}}
{2E}L$ in the oscillation probabilities to measure true CP
violation effects.

\section{T-odd asymmetries}
Since the matter contribution to the effective neutrino and 
antineutrino Hamiltonians is real and the matter density is symmetric
along the path of the neutrino beam in terrestrial LBL-experiments,
matter effects are T-symmetric. A non-vanishing value of $T_{\alpha
\beta}$ or $\bar{T}_{\alpha \beta}$, Eq.~(\ref{eq:diez}), in matter
can only be due to a fundamental violation of T-invariance. It is 
straightforward to obtain these T-odd asymmetries

\begin{equation}
T_{\alpha \beta} = \sum_{k>j} I'_{\alpha \beta; jk} \sin
\frac{\Delta \tilde{m}^2_{\nu k j }}{2E} L \quad ; \quad
\bar{T}_{\alpha \beta} = \sum_{k>j} \bar{I}'_{\alpha \beta, jk}
\sin \frac{\Delta \tilde{m}^2_{\bar{\nu} k j}}{2E} L
\label{eq:veintisiete}
\end{equation}

One needs, however, the joint measurements of $\nu_\mu \rightarrow
\nu_e$ and $\nu_e \rightarrow \nu_\mu$ which probably has to wait
for neutrino factories in muon-storage-ring facilities.

\section{Majorana Neutrinos}

If neutrinos are Majorana particles, the main modification for charged
current neutrino interaction is that the mixing matrix gets  replaced
\cite{ONCE} by

\begin{equation}
U^D \rightarrow U^M = U^D P \quad ; \quad  P = \left( \begin{array}{ccc}
e^{i \alpha_1} &&\\
& e^{i \alpha_2} &\\
&&  1 \end{array} \right)
\label{eq:veintiocho}
\end{equation}

\noindent
with $U^D$ the conventional $U$ complex mixing matrix with 3 mixing angles
and 1 CP-phase. One sees that 2 additional CP-phases come into the game. 
Some care is needed because CP-violation  through these new phases
means $\alpha_k \neq 0, \frac{\pi}{2}$!

The new phases are, however, not operative as long as we consider Dirac
flavour oscillations for neutrinos propagating through the Green
function \\ $<0|T \{\psi (x) \bar{\psi} (0) \} | 0>$. Their manifestation
would need the study of "neutrino-antineutrino" propagation mediated
by the Green function $< 0|T \{\psi (x) \psi^T (0) \} | 0>$.
  Langacker et al. \cite{DOCE} have shown that this conclusion
is valid not only for neutrino oscillations in vacuum but for
flavour oscillations in matter too. The new phases associated with
Majorana  neutrinos are not seen in Flavour Oscillations in Matter.

\section{Outlook}

The responses to the questions posed in the Introduction are
summarized now:

- CP,T violation in neutrino oscillation is possible in Appearance
Experiments, for mixing of 3 or more non-degenerate neutrinos.

- Non-vanishing  CP, T violating observables need the active
participation of the 3 (different) masses, mixings and CP-phase.
In particular, $\frac{E}{L}$ should be comparable to the smallest
$\Delta m^2$.

- In LBL experiments, there are large  matter effects, inducing fake
CP and CPT violation. The identification of true CP violation  would
need the explicit separation of odd functions of time.

-Matter Effects are T-symmetric, so that a non-vanishing T-odd
asymmetry in matter is still a signal of fundamental T-violation.

- If neutrinos are Majorana particles, there is no change for
Flavour Oscillations, either in vacuum or in matter.

\section*{Acknowledgments}
This work has been supported by the Spanish CICYT, under
Grant AEN-96/1718.


\begin{thebibliography}{99}
\bibitem{UNO} N. Cabiblo, \Journal{\PLB}{72}{333}{1978}.

\bibitem{DOS} S. M. Bilenky, J. Hosek, S.T.
Petcov, \Journal{\PLB}{94}{495}{1980}

\bibitem{TRES} M.Tanimoto,   \Journal{\PRD}{55}{322}{1997}.

\bibitem{CUATRO} J. Arafune, J. Sato,  \Journal{\PRD}{55}{1653}{1997};
J. Arafune, M. Koike, J. Sato, \Journal{\PRD}{56}{3093}{1997}.

\bibitem{CINCO} H. Minakata, H. Hunokawa, \Journal{\PLB}{413}{369}{1997};
\Journal{\PRD}{57}{4403}{1998}.

\bibitem{SEIS} S. M. Bilenky, C. Giunti, W. Grimus,
 \Journal{\PRD}{58}{033001}{1998}.

\bibitem{SIETE} A. De R\'ujula, M.B. Gavela, P. Hern\'andez,
CERN-TH/98-321.
\bibitem{OCHO} T.K. Kuo, J. Pantaleone, \Journal{\PLB}{198}{406}{1987}.
\bibitem{NUEVE} P.I. Krastev, S.T. Petcov, \Journal{\PLB}{205}{84}{1988}.
\bibitem{DIEZ} S. Toshev, \Journal{\PLB}{226}{335}{1989}.
\bibitem{ONCE} J. Bernab\'eu, P. Pascual, \Journal{\NPB}{228}{21}{1983}.
\bibitem{DOCE} P. Langacker, S. T. Petcov, G. Steigman, S. Toshev,
\Journal{\NPB}{282}{589}{1987}.


\end{thebibliography}
\end{document}